\documentclass[sigconf, nonacm]{acmart}
\usepackage{graphicx}
\usepackage{color}
\usepackage{balance}  
\usepackage{tabularx}
\usepackage{caption}
\usepackage{amsthm}
\usepackage{algorithm2e}
\usepackage{booktabs}
\usepackage{multirow}
\usepackage{caption}
\usepackage{subcaption}
\usepackage{enumitem}
\usepackage[a-1b]{pdfx}

\theoremstyle{definition} 
\newtheorem{exmp}{Example NL Query}
\newtheorem{chll}{Challenge}

\newcommand\vldbdoi{10.14778/3446095.3446103}
\newcommand\vldbpages{XXX-XXX}
\newcommand\vldbvolume{14}
\newcommand\vldbissue{5}
\newcommand\vldbyear{2021}
\newcommand\vldbauthors{\authors}
\newcommand\vldbtitle{\shorttitle}

\newcommand\vldbavailabilityurl{https://github.com/arifusta/DBTagger}
\newcommand\vldbpagestyle{empty} 

\begin{document}
\title{DBTagger:  Multi-Task Learning for Keyword Mapping in NLIDBs Using Bi-Directional Recurrent Neural Networks
\\\normalsize To appear in VLDB 2021}

\author{Arif Usta}
\affiliation{%
  \institution{Bilkent University}
  \city{Ankara}
  \state{Turkey}
}
\email{arif.usta@bilkent.edu.tr}

\author{Akifhan Karakayali}
\orcid{0000-0002-1825-0097}
\affiliation{%
  \institution{Bilkent University}
  \city{Ankara}
  \state{Turkey}
}
\email{akifhan@bilkent.edu.tr}

\author{\"{O}zg\"{u}r Ulusoy}
\orcid{0000-0001-5109-3700}
\affiliation{%
  \institution{Bilkent University}
  \city{Ankara}
  \state{Turkey}
}
\email{oulusoy@cs.bilkent.edu.tr}
\begin{abstract}
Translating Natural Language Queries (NLQs) to Structured Query Language (SQL) in interfaces deployed in relational databases is a challenging task, which has been widely studied in database community recently. Conventional rule based systems utilize series of solutions as a pipeline to deal with each step of this task, namely stop word filtering, tokenization, stemming/lemmatization, parsing, tagging, and translation. Recent works have mostly focused on the translation step overlooking the earlier steps by using ad\-hoc solutions. In the pipeline, one of the most critical and challenging problems is keyword mapping; constructing a mapping between tokens in the query and relational database elements (tables, attributes, values, etc.). We define the keyword mapping problem as a sequence tagging problem, and propose a novel deep learning based supervised approach that utilizes POS tags of NLQs. Our proposed approach, called \textit{DBTagger} (DataBase Tagger), is an end-to-end and schema independent solution, which makes it practical for various relational databases. 
We evaluate our approach on eight different datasets, and report new state-of-the-art accuracy results, $92.4\%$ on the average. Our results also indicate that DBTagger is faster than its counterparts up to $10000$ times and scalable for bigger databases.
\end{abstract}

\maketitle

\pagestyle{\vldbpagestyle}
\begingroup\small\noindent\raggedright\textbf{PVLDB Reference Format:}\\
\vldbauthors. \vldbtitle. PVLDB, \vldbvolume(\vldbissue): \vldbpages, \vldbyear.\\
\href{https://doi.org/\vldbdoi}{doi:\vldbdoi}
\endgroup
\begingroup
\renewcommand\thefootnote{}\footnote{\noindent
This work is licensed under the Creative Commons BY-NC-ND 4.0 International License. Visit \url{https://creativecommons.org/licenses/by-nc-nd/4.0/} to view a copy of this license. For any use beyond those covered by this license, obtain permission by emailing \href{mailto:info@vldb.org}{info@vldb.org}. Copyright is held by the owner/author(s). Publication rights licensed to the VLDB Endowment. \\
\raggedright Proceedings of the VLDB Endowment, Vol. \vldbvolume, No. \vldbissue\ %
ISSN 2150-8097. \\
\href{https://doi.org/\vldbdoi}{doi:\vldbdoi} \\
}\addtocounter{footnote}{-1}\endgroup

\ifdefempty{\vldbavailabilityurl}{}{
\vspace{.3cm}
\begingroup\small\noindent\raggedright\textbf{PVLDB Artifact Availability:}\\
The source code, data, and/or other artifacts have been made available at \url{\vldbavailabilityurl}.
\endgroup
}

\section{Introduction}

Amount of processed data has been growing rapidly pertaining to technology, leading database systems to have a great deal of importance in today's world. Amongst the systems, relational databases are still one of the most popular infrastructures to effectively store data in a structured fashion. To extract data out of a relational database, \textit{structured query language (SQL)} is used as a standard tool. Although SQL is a powerfully expressive language, even technically skilled users have difficulties using SQL. Along with the syntax of SQL, one has to know the schema underlying the database upon which the query is issued, which further causes hurdles to use SQL. Consequently, casual users find it even more difficult to express their information need, which makes SQL less desirable. To remove this barrier, an ideal solution is to provide a search engine like interface, such as Google or Bing in databases. The goal of \textit{Natural Language Interfaces to Databases (NLIDB)} is to break through these barriers to make it possible for casual users to employ their natural language to extract information.

To this end, many works have been published recently attacking the research problem of translation of natural language queries into SQL; such as conventional pipeline based approaches \cite{SODA, NALIR, ATHENA, Sqlizer} or end-to-end solutions using encoder-decoder based deep learning approaches \cite{iyer-etal-2017-learning, zhong2017seq2sql, xu2017sqlnet, syntaxSQL, dbpal2020}. Neural network based solutions seem promising in terms of robustness, covering semantic variations of queries. However, they struggle for queries requiring translation of complex SQL queries, such as aggregation and nested queries, especially if they include multiple tables. They also have a huge drawback in that they need many SQL-NL pairs for training to perform well, which makes conventional pipeline based solutions still an attractive alternative. \cite{challenges2020}. 

In the translation pipeline, one of the most important sub-problems is \textit{keyword mapping}, as noted in \cite{keywordChallenge} as an open challenge to be addressed in NLIDBs. \textit{Keyword mapping} task requires to associate each token or a series of consecutive tokens (e.g., keywords) in the natural language query to a corresponding database schema element such as table, attribute or value. It is the very first step of resolving ambiguity for translation. Xu et. al \cite{xu2017sqlnet} also note that during the translation of the query, \textit{where} clause is the most difficult part to generate which further signifies the task of \textit{keyword mapping}.

\begin{figure}
  \setlength\belowcaptionskip{-0.7\baselineskip}
  \includegraphics[width=0.95\linewidth]{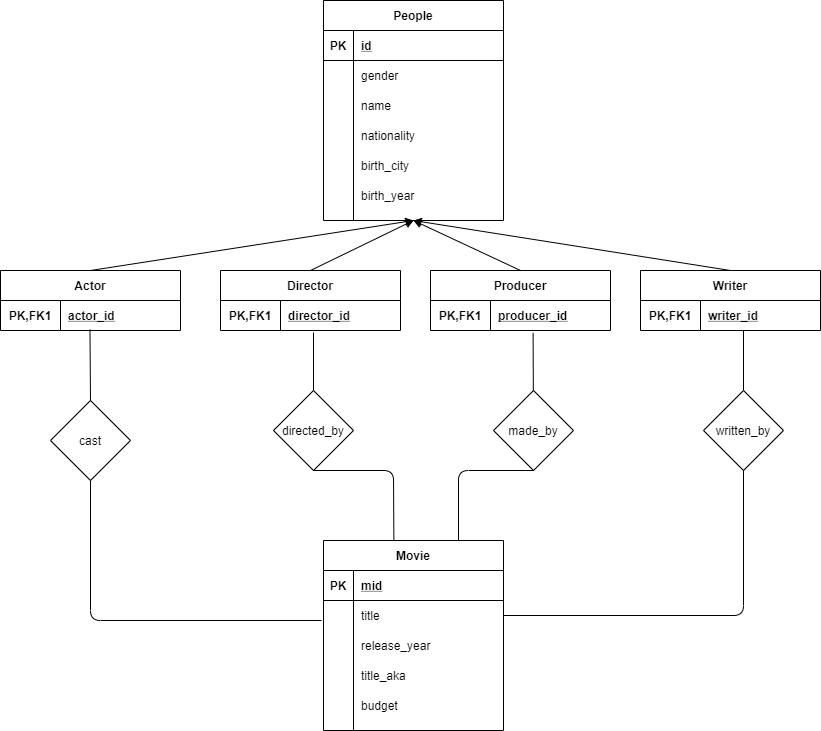}
  \caption{ER diagram of a subset of IMDB movie database}
  \label{fig:imdb_ER}
\end{figure}

Consider the below natural language query examples run on the sample IMDB movie database shown in Figure \ref{fig:imdb_ER} to better understand the challenges in \textit{keyword mapping} problem.

\begin{exmp}
\textit{"What is the writer of The Truman Show?"}
\end{exmp}

\begin{chll}
The very first challenge in \textit{keyword mapping} is to differentiate and categorize tokens in the query either as database relevant or not. For instance, some of the words in Example 1 (e.g., "is", "the", "of") are just stop words that are needed not to be considered as potential mapping target. An ad-hoc solution is to filter certain words using a pre-defined vocabulary, however such a solution removes "The" in Example 1 preceding the actual database value that needs to be mapped, which will cause the wrong translation. 
\end{chll}

\begin{chll}
Another important challenge is to detect multi-word entities (mostly database values), "The Truman Show" in  Example 1. The most common approach is to build look-up tables or indexes on n-grams of database values and calculate semantic and/or lexical similarity over the candidates. Yet, this is a costly process for on-the-fly calculations regarding possible n-grams of the qiven NL query. 
\end{chll}

\begin{exmp}
\textit{"Find all movies written by Matt Demon."}
\end{exmp}

\begin{exmp}
\textit{"How many movies are there that are directed by Steven Spielberg and featuring Matt Demon?"}
\end{exmp}

\begin{chll}
Consider the queries given in Examples 2 and 3. In the queries, tokens ("written" and "featuring") referring to database tables are syntactic and semantic variations of the actual table ("written\_by" and "cast" respectively) that they mapped to in the database (Figure \ref{fig:imdb_ER}). To handle such a challenge, lexical and semantic similarities of tokens over database elements (table and attributes) can be calculated using a third party database such as WordNet \cite{wordNet}. However, in addition to being a costly process to calculate such similarities online, such a solution cannot cover all possible variations of every map target in the database schema. Also, similarity calculation approach requires a manually crafted threshold, $\tau$, to determine how much similarity is sufficient to map to a particular schema element, which makes it undesirable.
\end{chll}

\begin{chll}
One of the usages of \textit{keyword mapping} step is to resolve ambiguities before getting into translation step. In the above examples, "Matt Demon" refers to a database value residing in multiple tables (e.g., actor, writer). Actual mapping of the keyword is determined by the mappings of neighbouring words surrounding, which implies that query-wise labelling considering \textit{coherence} rather than independent labelling can be beneficiary.
\end{chll}

\begin{chll}
In addition to an effective solution, an ideal keyword mapping approach must be efficient to be deployed on interfaces where users run queries online. Mapper should output the result in reasonable time.
\end{chll}

Most of the pipeline-based state-of-the-art works do not provide a novel solution to the problem of \textit{keyword mapping}, rather they utilize \textcolor{black}{unsupervised approaches such as} simple look-up tables looking for exact matches or pre-defined synonyms \cite{PrecisePopescu, SODA}; or they make use of an existing lexical database \cite{NALIR} such as WordNet \cite{wordNet}; or they exploit domain information to extract an ontology to be used for the task \cite{ATHENA}; or they employ distributed representations of words \cite{Sqlizer} such as word2vec \cite{word2vec} to calculate semantic similarity of tokens over database elements. Although these approaches are effective to some extent, they fail to solve various challenges mentioned by the task of \textit{keyword mapping} single-handedly.

In order to address all of the challenges mentioned above we propose DBTagger, a novel deep sequence tagger architecture used for \textit{keyword mapping} in NLIDBs. Our approach is applicable to different database domains requiring only handful of training query annotations and practical to be deployed in online scenarios finding tags in just milliseconds. In particular, we make the following contributions by proposing DBTagger:

\begin{figure*}[t!]
    \centering
    \includegraphics[width=\textwidth, height=11cm,keepaspectratio]{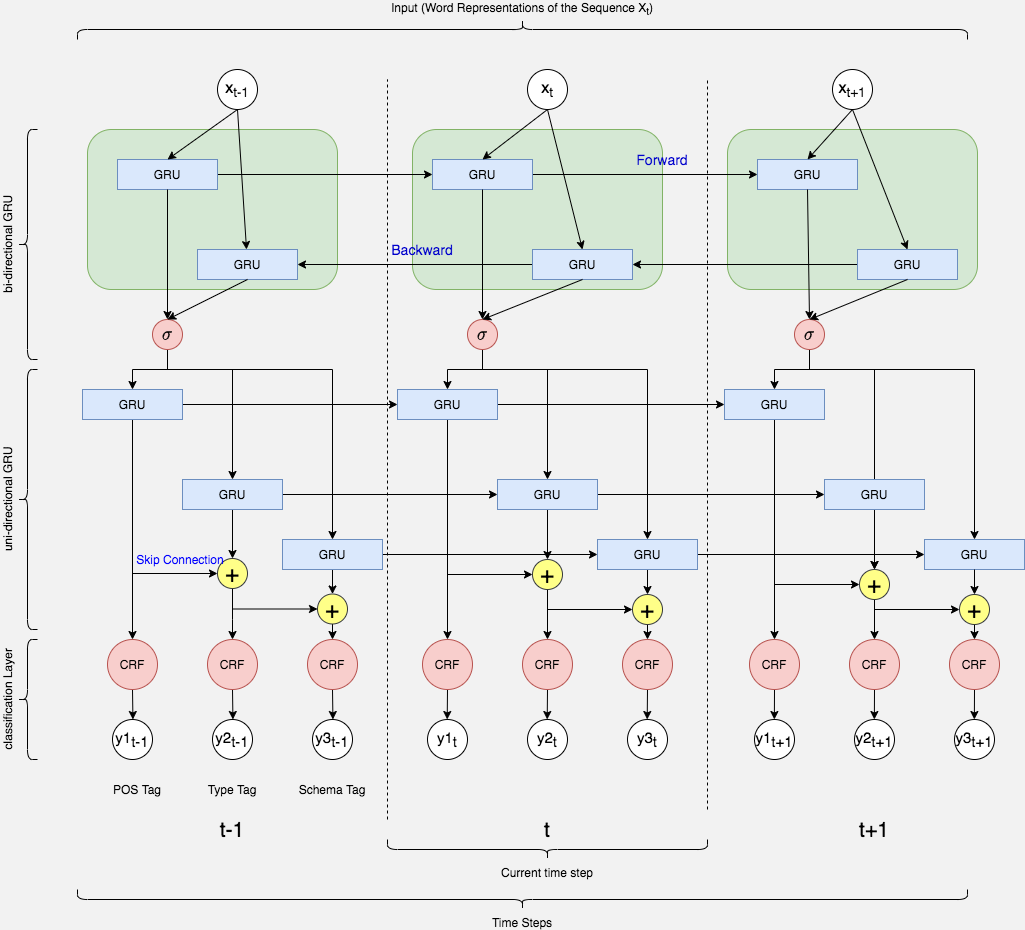}
    \caption{\textcolor{black}{DBTagger Network}}
    \label{fig:NNStructure2}
\end{figure*}

\begin{itemize}
    \item We tackle the keyword mapping problem as a sequence tagging problem and borrow state-of-the-art deep learning approaches tailored for well-known NLP tasks. 
    \item We extend the neural structure for sequence tagging, 
    \textcolor{black}{by utilizing \textit{multi-task learning} and \textit{cross-skip connections}} to exploit the observation we made in natural language query logs of databases, that is, schema tags of keywords are highly correlated with POS tags.
    \item We manually annotate 
    query logs from three publicly available relational databases, and \textcolor{black}{five different schemas belonging to Spider \cite{yu-etal-2018-spider} dataset}.
    \item We evaluate DBTagger, with above-mentioned query logs in two different setups. \textcolor{black}{First, we compare DBTagger with unsupervised baselines preferred in state-of-the-art NLIDBs.} \textcolor{black}{In the latter, we evaluate DBTagger architecture by comparing with different supervised neural architectures. We report new state-of-the-art accuracy results for keyword mapping in all datasets.}
    \item \textcolor{black}{We provide comprehensive run time and memory usage analysis over the existing keyword mapping approaches. Our results show that, DBTagger is the most efficient and scalable approach for both metrics.}  
\end{itemize}

The remainder of this paper is organized as follows. In the next section, we explain the problem formulation and methodology we follow. We present the neural network structure we designed to solve the keyword mapping problem, and discuss how annotation of queries is handled. 
In Section 3, we provide experimental results comparing DBTagger \textcolor{black}{with unsupervised baselines} and \textcolor{black}{present the performance of different neural models to justify DBTagger architecture.} \textcolor{black}{We also provide an efficiency analysis on all baselines.} In Section 4, we summarize related work.
 We conclude with a discussion of our results and outline possible future work plans in Section 5.

\section{Methodology}


\subsection{Deep Sequence Tagger Architecture}
\textcolor{black}{POS tagging and NER refer to sequence tagging problem in NLP for a particular sentence to identify parts-of-speech such as noun, verb, adjective and to locate any entity names such as person, organization, respectively.} 
Recurrent Neural Networks (RNN) are at the core of architectures to handle such problems, since they are a family of networks that perform well on sequential data input such as a sentence. 

\textcolor{black}{In RNN networks, the basic goal is to carry past information (previous words) to future time steps (future words) to determine values of inner states and consequently the final output, which makes them preferable architecture for sequential data. Given $x_t$ as input at time step $t$, calculation of hidden state $h_t$ at time step $t$ is as follows:} 

\begin{equation}
    h_t = f(Ux_t + Wh_{t-1})
\end{equation}

\textcolor{black}{In practice, however, RNN networks suffer from \textit{vanishing gradient problem}, therefore the limitation was overcome by modifying the gated units of RNNs; such as LSTM \cite{LSTM} and GRU\cite{GRU}. Compared to vanilla RNN, LSTM has \textit{forget gates} and GRU comprises of \textit{reset} and \textit{update} gates additionally. 
We experimented with both structures and we chose GRU due to its better performance in our experiments. In GRU, Update Gates decide what information to throw away and what new information to add, whereas Reset Gate is utilized to decide how much past information to forget.} 

In sequence tagging problem, in addition to past information we also have future information as well at a given specific time, $t$. \textcolor{black}{For a particular word $w_i$, we know the preceding words (past information) and succeeding words (future information), which can be further exploited in the particular network architecture called, \textit{bi-directional RNN} introduced in \cite{bi-directionalRNN}.} Bi-directional RNN has two sets of networks with different parameters called forward and backward. The concatenation of the two networks is then fed into the last layer, where the output is determined. This process is demonstrated in the upper part of the Figure \ref{fig:NNStructure2}, named bi-directional GRU.

Sequence tagging is a supervised classification problem where the model tries to predict the most probable label from the output space. For that purpose, although conventional \textit{softmax} classification can be used, \textit{conditional random field (CRF)} \cite{CRF} is preferred. Unlike independent classification by softmax, CRF tries to predict labels sentence-wise by taking labels of the neighboring words into consideration as well. This feature of CRF is what makes it an attractive choice especially in a problem like \textit{keyword mapping}. CRFs for each class of tags are appended to uni-directional GRU, depicted in lower part of the Figure \ref{fig:NNStructure2}.


\subsection{DBTagger Architecture}

Formally, for a given NL query, input $X$ becomes a series of vectors $[x_1,x_2,...x_n]$ where $x_i$ represents the $i^{th}$ word in the query. Similarly, output vector $Y$ becomes $[y_1,y_2,...y_n]$ where $y_i$ represents the label (actual tag) of the $y^{th}$ word in the query. Input must be in numerical format, which implies that a numerical representation of words is needed. For that purpose, the word embedding approach is state-of-the-art in various sequence tagging tasks in NLP \cite{collobert2011} before feeding into the network. So, embedding matrix is extracted for the given query, $W\in R^{nxd} $, where $n$ is the number of words in the query and $d$ is the dimension of the embedding vector for each word. 
For the pre-calculated embeddings, we used fastText\cite{fasttext} due to it being one of the representation techniques considering sub-word (character n-grams) as well to deal with the out of vocabulary token problem better.

\textcolor{black}{
We consider $G$ to be 2-dimensional scores of output by the uni-directional GRU with size $n\times k$ where $k$ represents the total number of tags. $G_{i,j}$ refers to score of the $j^{th}$ tag for the $i^{th}$ word. For a sequence $Y$ and given input $X$, we define tag scores as;
}

\begin{equation}
    s(X,Y) = \sum_{i=1}^{n}A_{y_i,y_{i+1}} + \sum_{i=1}^{n}G_{i,y_i}
\end{equation}

\textcolor{black}{where $A$ is a transition matrix in which $A_{i,j}$ represents the score of a transition from the $i^{th}$ tag to the $j^{th}$ tag. After finding scores, we define probability of the sequence $Y$:
}

\begin{equation}
    p(Y|X) = \frac{e^{s(X,Y)}} { \sum_{\bar{Y}\in Y_x}^{}e^{s(X,\bar{Y})}}
\end{equation}

\textcolor{black}{where $\bar{Y}$ refers to any possible tag sequence. During training we maximize the log-probability of the correct tag sequence and for the inference we simply select the tag sequence with the maximum score.}

\textcolor{black}{In our architecture, we utilize \textit{Multi-task learning} by introducing two other related tasks; POS and type levels (shown in Figure \ref{fig:NNStructure2}). The reason we apply multi-task learning is to try to exploit the observation that actual database tags of the tokens in the query are related to POS tags. Besides, multi-task learning helps to increase model accuracy and efficiency by making more generalized models with the help of shared representations between tasks \cite{caruana1997multitask}. POS and Type tasks are trained with schema task to improve accuracy of schema (final) tags.  For each task, we define the same loss function, described above. During backpropagation, we simply combine the losses as follows;
}

\begin{equation}
    \begin{split}
    L_{total} &= \sum_{i=1}^{3}w_i\times L_i \textrm{ subject to}\\
    \sum_{i=1}^{3}w_i &= 1 
\end{split}
\end{equation}

where $w_i$ represents the weight of $i^{th}$ task and $L_i$ represents the loss calculated for the $i^{th}$ task similarly.  



Another technique we integrate into the neural architecture is \textit{skip-connection}. Skip connection is used to introduce extra node connections between different layers by skipping one or more layers in the architecture.\textcolor{black}{With skip connections, the model provides an alternative for gradient to back propagation, which eventually helps in convergence.} The technique has become \textcolor{black}{compulsory} component in many neural architectures deployed in computer vision community, such as the famous architectures ResNet \cite{ResNet} and DenseNet \cite{DenseNet}. In the architecture of DBTagger, for each task except the first one (POS), we additionally feed the output of uni-directional GRU layer of previous task into CRF layer of the next task (${i+1}^{th}$ task). With these connections, we further carry the information of previous tasks to later tasks and eventually to the final task, schema tagging. 

\begin{table*}[t]
    \centering
    \caption{An example NL query with its tags corresponding to each word in three different levels}
    \begin{tabular}{l|llllllllll}
    \toprule
         \textbf{NL query}& who &acted &John &Nash &in &the &movie &A &Beautiful &Mind \\
         \\[-1em]
         \textbf{POS tags}& WP & VBD & NNP & NNP & IN & DT & NN & DT & JJ & NN\\
         \\[-1em]
         \textbf{Type tags}& O & TABLEREF & VALUE & VALUE & COND & O & TABLE & VALUE & VALUE & VALUE \\
         \\[-1em]
         \textbf{Schema tags}& O & cast & cast.role & cast.role & cond & O & movie & movie.title & movie.title & movie.title\\
    \bottomrule
    \end{tabular}
    \label{tab:ex_tagging}
\end{table*}


\subsection{Annotation Scheme}

In our problem formulation, every token (words in the natural language query) associates three different tags; namely part-of-speech (POS) tag, type tag and schema tag. In the following subsections, we explain \textcolor{black}{how we extract or annotate} each of them in detail.
\subsubsection{POS Tags}
To obtain the POS tags of our natural language queries we used the toolkit of Stanford Natural Language Processing Group named Stanford CoreNLP\cite{coreNLP}. We use them as they are output from the toolkit, without doing any further processing \textcolor{black}{since the reported accuracy for POS Tagger (97\%) is sufficient enough.}   

\subsubsection{Type Tags}
In each natural language query, there are keywords (words or consecutive words) which can be mapped to database schema elements such as table, attribute or value. We divide this mapping into two levels; type tagging and schema tagging. Type tags represent the type of the mapped schema element to be used in the SQL query. In total we have seven different type tags;
\begin{itemize}[leftmargin=*]
    \item \textbf{TABLE}: NLQs contain nouns which may inhibit direct references to the tables in the schema, and we tag such nouns with \textit{TABLE} tag. In the example NL query given in Table \ref{tab:ex_tagging}, noun \textit{movie} has a type tag as TABLE, which also supports the intuition that schema labels and pos tags are related.
    \item \textbf{TABLEREF}: Although the primary sources for table references are nouns, some verbs contain references to the tables most of which are relation tables. TABLEREF tag is used to identify such verbs. Revisiting the example given Table \ref{tab:ex_tagging}, the verb \textit{acted} refers to the table \textit{cast}, and therefore it is tagged with TABLEREF to differentiate better the roles of POS tags in the query.
    \item \textbf{ATTR}: In SQL queries, attributes are mostly used in SELECT, WHERE and GROUP BY clauses. Natural language queries may contain nouns that can be mapped to those attributes. We use ATTR tag for tagging such nouns in the natural language queries.
    \item \textbf{ATTRREF}: Like TABLEREF tag, ATTRREF tag is used to tag the verbs in the natural language query that can be mapped to the attributes in the SQL query.
    \item \textbf{VALUE}: In NLQs, there are many entity like keywords that need to be mapped to their corresponding database values. These words are mostly tagged as \textit{Proper noun-NNP} such as the keyword \textit{John Nash} in the example query. In addition to these tags, it is also likely for a word to have a \textit{noun-NN} POS tag with a \textit{Value} tag corresponding to schema level. In order to handle these cases having different POS tags, we have \textit{Value} type tags (e.g., \textit{Mind} keyword in the example query is part of a keyword that needs to be mapped as \textit{value} to \textit{movie.title}). Keywords with \textit{Value} tags can later be used in the translation to determine "where" clauses in SQL.
    \item \textbf{COND}: After determining which keywords in the query are to be mapped as values, it is also important to identify the words that imply which type of conditions to be met for the SQL query. For that purpose, we have the \textit{COND} type tag.
    \item \textbf{O (OTHER)}: This type of tag represents words in the query that are not needed to be mapped to any schema instrument related to the translation step. Most stop words in the query (e.g., the) fall into this category. 
\end{itemize}

\begin{table*}[]
    \centering
    \caption{\textcolor{black}{Statistics of the databases used}}
    \begin{tabular}{lcccccccc}
    \toprule
         & \multicolumn{3}{c}{\textbf{Database}} & \multicolumn{5}{c}{\textbf{Spider}}\\
         \cmidrule(lr){2-9}
         Properties (\#)& imdb & mas & yelp 
         & academic & college & hr & imdb & yelp\\
         \midrule
         entity tables&6&7&2&7&5&6&6&2 \\
         relation tables&11&5&5&8&2&1&11&5 \\
         total tables&17&12&7&15&7&7&17&7 \\
         total attributes&55&28&38&42&43&35&55&38 \\
         nonPK-FK attributes&14&7&16&18&29&21&14&16 \\
         total tags&31&19&20&26&36&30&31&20 \\
         queries&131&599&128&181&164&124&109&110 \\
         tokens in queries&1250&4483&1234&2127&2130&2099&1012&1035 \\
         \bottomrule
    \end{tabular}
    \label{tab:comprops}
\end{table*}

\begin{table}[t]
    \centering
    \caption{\textcolor{black}{Accuracy scores of unsupervised baselines for relation and non-relation matching}}
    \begin{tabular}{lccc}
    \toprule
         & \multicolumn{3}{c}{\textbf{Database}}\\
         \cmidrule(lr){2-4}
         Baseline & imdb & mas & yelp 
         \\
         \midrule
         tf-idf&0.594-0.051&0.734-0.084&0.659-0.557\\
         NALIR&0.574-0.103&0.742-0.476&0.661-0.188\\
         word2vec&0.625-0.093&0.275-0.379&0.677-0.269\\
         TaBERT&NA-0.251&NA-0.094&NA-0.114\\
         \midrule
         DBTagger&0.908-0.861&0.964-0.950&0.947-0.923\\
         \bottomrule
    \end{tabular}
    \label{tab:unsupervisedScores}
    \vspace{-4mm}
\end{table}

\subsubsection{Schema Tag}
Schema tags of keywords represent the database mapping that the keyword is referring to; name of a table, or attribute. Tagging a keyword with a type tag is important yet incomplete. To find the exact mapping the keyword refers to, we defined a second level tagging where the output is the name of the tables or attributes. For each entity table (e.g. \textit{movie} table in Figure \ref{fig:imdb_ER}) and for each non-PK or non-FK attribute (attributes which have semantics) we define a schema tag (e.g \textit{movie, people, movie.title}, etc., referring to Figure \ref{fig:imdb_ER}). We complete possible schema tags by carrying \textit{OTHER} and \textit{COND} from type tags. 
We use the same schema tag for attributes and values (e.g \textit{movie.title}), but differentiate them at the inference step by combining tags from both type tags and schema tags. If a word is mapped into \textit{Value} type tag as a result of the model, its schema tag refers to the attribute in which the value resides.

In order to annotate queries, we annotate each word in the query for three different levels mentioned above. While POS tags are extracted automatically, we manually annotate the other two levels. \textcolor{black}{Annotations were done by three graduate and three undergraduate computer science students who are familiar with database subject. Although annotation time varies depending on the person, on the average it took a week to annotate tokens by a single person for two levels (type and schema) for a query log with $150$ NL questions, which we believe is practical to apply in many domains}.

\section{Experimental Evaluation}

\subsection{Datasets}

In our experiments we used \textcolor{black}{\textit{yelp, imdb \cite{Sqlizer}}, and \textit{mas \cite{NALIR}} datasets which are heavily used in many NLIDB related works by the database community \cite{NALIR, ATHENA, Sqlizer,jagadishBridge}. In addition to these datasets, we also used different schemas from the \textit{Spider} dataset \cite{yu-etal-2018-spider}; which are \textit{academic, college, hr, imdb}, and \textit{yelp}. Spider is comprised of approximately $200$ schemas from different domains; however, there are only handful (around 10) of schemas with more than $100$ NL questions. Number of questions is important for our deep learning based solution, since it requires certain number of training data to effectively train. Each schema we picked from Spider dataset is among the schemas with most number of NL questions, having over $100$ queries to work with. Due to the lack of sufficient database values (many schemas do not have database rows or have few number of rows), we used the Spider dataset only on supervised setup.}

\textcolor{black}{The statistics about each dataset for which annotation is done is shown in Table \ref{tab:comprops}. In Table \ref{tab:comprops} (referring to Figure \ref{fig:imdb_ER}), entity tables refer to main tables (i.e. Movie), relation tables refer to hub tables that store connections between entity tables (i.e. cast, written\_by), nonPK-FK attributes refer to attributes in any table that is neither PK nor FK (i.e., gender in People table), and finally total tags refer to unique number of taggings extracted from that particular schema depending on the above mentioned values. Final schema tags of a particular database are determined by composing table names and name of the nonPK-FK attributes in addition to COND and OTHER. In the last two rows of the Table \ref{tab:comprops}, we show annotated number of NL questions, referred to as queries, and the number of total words inside these queries, referred to as tokens.}

\subsection{Settings}

\textcolor{black}{We first split the datasets into train-validation sets with $5-1$ ratio, respectively to be used for tuning task weights.} \textcolor{black}{For models trained on multiple tasks, we used $0.1-0.2-0.7$ as tuned weights for POS, Type and Schema tasks, respectively.}

We train our \textcolor{black}{deep neural} models using the backpropagation algorithm with two different optimizers; namely Adadelta \cite{Zeiler2012ADADELTAAA} and Nadam \cite{nadam}. We start the training with Adadelta and continue it with Nadam. We found that using two different optimizers resulted better in our problem. For both shared and unshared bi-directional GRUs, we use $100$ units and apply dropout \cite{dropout} with the value of $0.5$ including recurrent inner states as well. For training, the batch-size is set to $32$ for all datasets. 
Parameter values chosen are similar to that reported in the study  \cite{lample-etal-2016-neural} (the state-of-the-art NER solution utilizing deep neural networks), such as the dropout and batch size values. \textcolor{black}{We measure the performance of each neural model by applying cross validation with 6-folds. All the results reported are the average test scores of 6-folds. During inference, we discard POS and Type task results and only use Schema (final) tasks to measure scores.}

\subsection{Results}

\subsubsection{Comparison with Unsupervised Baselines}
\textcolor{black}{
We implemented the unsupervised approaches utilized in the state-of-the art NLIDB works for the keyword mapping task as baselines to compare with DBTagger.}
\textcolor{black}{
\begin{itemize}[leftmargin=*]
    \item[-] \textbf{tf-idf:} Similar to ATHENA \cite{ATHENA}, for each unique value present in the database, we first create an exact matching index, and then perform tf-idf for tokens in the NLQ. In case of matches to multiple columns, the column with the biggest tf value is chosen as matching. In order to handle multi word keywords, we use n-grams of tokens up to $n=3$. For relation matching, we used lexical similarity based on the Edit Distance algorithm.
    \item[-] \textbf{NALIR:} \textcolor{black}{NALIR \cite{NALIR} uses WordNet 
    for relation matching. 
    For non-relation matching,
    it utilizes regex or full text search queries over each database column whose type is text. In case of matches to multiple columns, the column which returns more rows as a result is chosen as matching. For fast retrieval, we limit the number of rows returned from the query to $2000$, as in the implementation of NALIR.}
    \item[-] \textbf{word2vec:} For each unique value present in the database, cosine similarity over tokens in the NLQ is applied to find mappings using pre-defined wor2vec embeddings. The matching with the highest similarity over a certain threshold is chosen.
    \item[-] \textbf{TaBERT:} TaBert \cite{tabert} is a transformer based encoder which generates dynamic word representations (unlike word2vec) using database content. The approach also generates column encoding for a given table, which makes it an applicable keyword mapper for non-relation matching.
    For a particular token, matching with maximum similarity over a certain threshold is chosen.
\end{itemize}
}
\textcolor{black}{We categorize the keyword mapping task as \textit{relation matching} and \textit{non-relation matching}. The former mapping refers to matching for table or column names and the latter refers to matching for database values.
}\textcolor{black}{\textcolor{black}{For fair comparison, we do not apply any pre or post processing over the NL queries or use external source of knowledge, such as a keyword parser or metadata extractor.} Results are shown in Table \ref{tab:unsupervisedScores}. Each pair of scores represents token wise accuracy for relation and non-relation matching.} 

\textcolor{black}{DBTagger outperforms unsupervised baselines in each dataset significantly, by up to $31\%$ and $65\%$ compared to best counterpart for relation and non-relation matching, respectively. For relation matching, results of all approaches are similar to each other except the word2vec method for the mas dataset. The main reason for such poor performance is that the mas dataset has column names such as \textit{venueName} for which word2vec cannot produce word representations, which radically reduces chances of semantic matching.} 

\textcolor{black}{tf-idf gives promising results on the yelp dataset, whereas it fails on the imdb and mas datasets for non-relation matching. This behavior is due to presence of ambiguous values (the same database value in multiple columns) and not being able to find a match for values having more than three words. For the \textit{imdb} dataset, none of the baselines performs well for non-relation matching. The \textit{imdb} dataset has entity like values that are comprised of multiple words such as movie names, which makes it impossible for semantic matching approaches to generate meaningful representations to perform similarity. NALIR's approach of querying over database has difficulties for the imdb and yelp datasets since the approach does not solve ambiguities without user interaction. TaBERT performs poorly for all datasets. TaBERT has its own tokenizer, which tries to deal with tokens that are out of vocabulary (OOV) by breaking the token into sub-words that have representations. OOV keywords appearing in the natural language query are therefore divided by the tokenizer into pieces, which eventually leads to unrelated word representations and therefore non-predictive similarity calculation.} 

\subsubsection{Translation Accuracy}
\textcolor{black}{In order to show the effectiveness of tags output by DBTagger, we implemented a simple translation pipeline, similar to methodology in \cite{jagadishBridge}. The pipeline generates join paths for SQL translation using shortest length path over schema graph to cover all the mappings output by DBTagger. We count inaccurate, if the algorithm can not output a joining path. We compare our pipeline with a state-of-the-art system, NALIR\cite{NALIR}, and TEMPLAR\cite{jagadishBridge}, which is an enhancer over an existing NLIDB system. The results are presented in Table \ref{tab:translationScores}. The pipeline over DBTagger tags outperforms both systems in imdb and mas datesets, up to $66\%$ and $37\%$ compared to NALIR and TEMPLAR respectively. For queries which do not include nested or group by constraints such as simple select-join queries, our pipeline produces $67\%$, $77\%$ and $53\%$ translation accuracy for imdb, mas and yelp datasets respectively. Considering the simplicity of the translation algorithm, results demonstrate the efficacy of predicted outputs of DBTagger.} 

\begin{table}[t]
    \centering
    \caption{\textcolor{black}{Translation Accuracy}}
    \begin{tabular}{lccc}
    \toprule
         & \multicolumn{3}{c}{\textbf{Database}}\\
         \cmidrule(lr){2-4}
         NLIDB System& imdb & mas & yelp \\
         \midrule
         NALIR &0.383& 0.330 &0.472\\
         TEMPLAR (on NALIR) &0.500&0.402 & 0.528\\
         \midrule
         DBTagger Pipeline & 0.564& 0.551& 0.461\\
         \bottomrule
    \end{tabular}
    \label{tab:translationScores}
        \vspace{-8mm}
\end{table}

\begin{table*}[]
    \centering
    \caption{\textcolor{black}{Performance of Neural Models with Different Architectures in accuracy-F1 metrics}}
    \begin{tabular}{lccc|ccccc}
    \toprule
         & \multicolumn{3}{c}{\textbf{Database}} & \multicolumn{5}{c}{\textbf{Spider}}\\
         \cmidrule(lr){2-9}
         Model & \multicolumn{1}{c}{yelp} & \multicolumn{1}{c}{imdb} & \multicolumn{1}{c}{mas} 
         & \multicolumn{1}{c}{academic} & \multicolumn{1}{c}{hr} & \multicolumn{1}{c}{college} & \multicolumn{1}{c}{imdb} & \multicolumn{1}{c}{yelp}\\
         \midrule
         CRF & 0.934-0.890&0.907-0.850&0.955-0.932 & \textbf{0.974}-\textbf{0.956} & \textbf{0.881}-\textbf{0.748}& 0.878-0.721& 0.866-0.821 &  0.880-0.827\\
         ST\_Uni &0.939-0.883 &0.905-0.805&0.961-0.938 & 0.962-0.945 &0.844-0.642 & 0.854-0.692&0.848-0.751 & 0.865-0.803 \\
         ST\_Bi &0.947-0.908&0.917-0.832& 0.964-0.941 & 0.966-0.952& 0.877-0.689& 0.872-0.720&0.882-0.811 & 0.891-0.841 \\
         MT\_Seq &0.938-0.886&0.921-0.853&0.964-\textbf{0.943}& 0.964-0.952& 0.835-0.685 & 0.886-0.714& 0.896-0.837 & 0.895-0.838 \\
         DBTagger &\textbf{0.968}-\textbf{0.938}&\textbf{0.935}-\textbf{0.878}&\textbf{0.965}-0.941& 0.965-0.954& 0.861-0.735 &\textbf{0.904}-\textbf{0.761} &\textbf{0.898}-\textbf{0.855} & \textbf{0.897}-\textbf{0.854} \\
         \bottomrule
    \end{tabular}
    \label{tab:evaluationArch}
\end{table*}

\begin{figure*}
     \centering
     \begin{subfigure}[b]{0.48\textwidth}
         \centering
         \includegraphics[height=6cm]{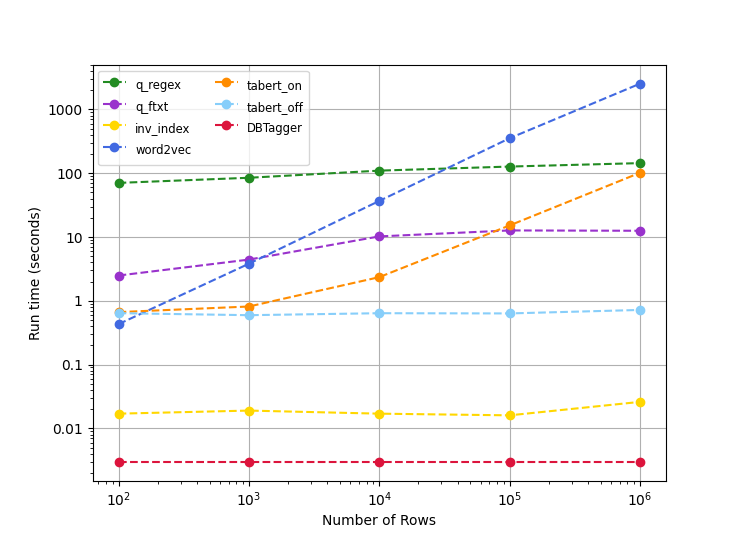}
         \caption{Run Time}
     \end{subfigure}
     \begin{subfigure}[b]{0.44\textwidth}
         \centering
         \includegraphics[height=6cm]{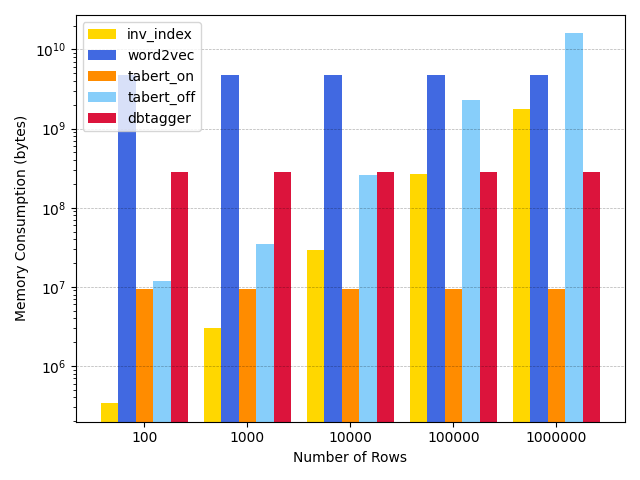}
         \caption{Memory Usage}
     \end{subfigure}
     \caption{\textcolor{black}{Run Time and Memory Usage of state-of-the-art keyword mapping approaches }}
     \label{fig:runtimes}
\end{figure*}

\subsection{Impact of DBTagger Architecture}
\textcolor{black}{In this experimental setup, we perform keyword mapping in a supervised fashion with different neural network architectures along with a non-Deep Learning (DL) baseline to evaluate architectural decisions.
}

\textcolor{black}{
\begin{itemize}[leftmargin=*,topsep=3pt]
    \item[-] \textbf{CRF:} As a non-DL baseline, we use vanilla CRF. Semantic word representations of the NLQ are fed as input to the model.
    \item[-] \textbf{ST\_Uni:} We create a two layers stack of uni-directional GRUs, followed by CRF as the classification layer. This model is trained on only a single task, schema tags.
    \item[-] \textbf{ST\_Bi:} Different than the previous architecture, we use bi-directional GRUs instead of uni-directional GRUs. Classification is done on the CRF layer.
    \item[-] \textbf{MT\_Seq} In this model, training is performed on all three tasks. However, each task is trained separately. The predicted tag of the previous task is fed into the next task. To do that, 1-hot vector representations of predicted tags are concatenated with semantic word representations. We stack a bi-directional GRU with a uni-directional GRU to encode the sentence and feed the output vector to the CRF layer.
    \item[-] \textbf{DBTagger} This model represents the DBTagger architecture where all tasks are used during training concurrently. DBTagger also has cross-skip connections between tasks as depicted in Figure \ref{fig:NNStructure2}.
\end{itemize}
}

\textcolor{black}{For all the models, the same hyper parameters are used for fair comparison during training, as explained in Section3.2. The results are shown in Table \ref{tab:evaluationArch}. Each pair of scores represents the accuracy and F1 measures, respectively.  DBTagger performs better than the other supervised architectures for six different datasets in accuracy and in terms of F1. Especially for the yelp and college datasets the performance improvement is remarkable, which is up to around $4.5\%$ and $5\%$, respectively. Vanilla CRF performs well among all (best in two datasets), which signifies its role in the architecture for the sequence tagging problem. ST\_Bi performs better than ST\_Uni in all datasets, which shows the positive impact of bi-directional GRUs. Compared to single task models, multi task models perform better for all datasets. Except the \textit{mas} dataset for the F1 metric, DBTagger produces better tags compared to the other multi task model, MT\_Seq, in which tasks are trained separately.}

\subsection{Efficiency Analysis}
\textcolor{black}{Efficiency is one of the most important properties of a good keyword mapper to be deployable in online interfaces. Therefore, run time performance of keyword mapping approaches mentioned in Section 3.3 is also evaluated.}

\textcolor{black}{
\begin{itemize}[leftmargin=*,topsep=0pt]
    \item[-] \textbf{NALIR}: We analyze both querying over database column approaches used in NALIR\cite{NALIR}, named as \textit{q\_regex} and \textit{q\_ftext}, which use \textit{like} and \textit{match against} operators respectively.
    \item[-] \textbf{tf-idf}: Similar to ATHENA \cite{ATHENA}, we created an exact matching index, using inverted index named as \textit{inv\_index}, beforehand to avoid querying over database, . 
    \item[-] \textbf{word2vec}: Many works such as Sqlizer \cite{Sqlizer} make use of pre-trained word embeddings to find mappings, which requires keeping the model in the memory to perform similarities.
    \item[-] \textbf{tabert\_on}: TaBert requires database content (content snapshot) to generate encodings for both NL tokens and columns. We call this setup tabert online, where the model generates the content snapshot to perform mapping when the query comes. 
    \item[-] \textbf{tabert\_off}: We also use TaBert in offline setup. For each table, database content is generated beforehand to perform encodings. In this setup, we keep the content in the memory to serve the query faster.
\end{itemize}
}

\textcolor{black}{We measured the time elapsed for a single query to extract tags and the memory consumption needed to perform mapping for each approach. We also run each experiment with different number of row values to capture the impact of the database size. Figure \ref{fig:runtimes} presents run time and memory usage analysis of keyword mappers. DBTagger ouputs the tags faster than any other baseline and it is scalable to much bigger databases. However, q\_regex, q\_ftext, tabert\_on and word2vec do not seem applicable for bigger tables having more than $10000$ rows. The tf-idf technique has nice balance between run-time and memory usage, but it is limited in terms of effectiveness (Table \ref{tab:unsupervisedScores}). tabert-off performs the tagging in a reasonable time, yet it requires huge memory consumption especially for bigger tables. 
}

\section{Related Work}

\subsection{NLIDBs and Keyword Mapping Approaches}
Although the very first effort \cite{firstNLIDB} of providing natural language interface in databases dates back to multiple decades ago, the popularity of the problem has increased due to some recent pipeline based systems proposed by the database community, such as SODA \cite{SODA}, NALIR \cite{NALIR}, ATHENA\cite{ATHENA} and SQLizer\cite{Sqlizer}. 

Recently, end-to-end approaches utilizing encoder-decoder based architectures \cite{zhong2017seq2sql, xu2017sqlnet, dbpal2018, yavuz-etal-2018-takes, yu-etal-2018-typesql, huang-etal-2018-natural, syntaxSQL, IRNet2019-towards, bogin-etal-2019-representing, dbpal2020} in deep learning have become more popular to deal with the translation problem. 
Seq2SQL\cite{zhong2017seq2sql}  uses a Bi-LSTM to encode a sequence that contains columns of the related table, SQL keywords and question. 
The study \cite{zhong2017seq2sql} also provided a dataset called \textit{WikiSql} to the research community working on NLIDB problem for evaluation. 
SQLNet\cite{xu2017sqlnet} defines a seq-to-set approach to eliminate reinforcement learning process of Seq2SQL. 
In another study which used WikiSql dataset, Yavuz et al.\cite{yavuz-etal-2018-takes} employs a process called candidate generation to create keyword mappings to be used in \textit{where} clasue in SQL translation specifically. 
All of the proposed deep learning based methods use pre-trained word embedding vectors for input to the model. Therefore, keyword mapping is implicitly handled by the model. However, TypeSql \cite{yu-etal-2018-typesql} tries to enrich input data augmenting entity tags by performing similarity check over the database or knowledge base. 
Similary, \cite{huang-etal-2018-natural} tries to find possible constant values in the query by performing similarity matching.

Due to the limited nature of WikiSql dataset, having a single table for each database, another important dataset called Spider \cite{yu-etal-2018-spider} is provided to the community. Consequently, many studies proposed recently \cite{syntaxSQL, IRNet2019-towards, bogin-etal-2019-representing, dbpal2020, tabert} have evaluated their solutions on the Spider dataset. \textcolor{black}{Different from the others, TaBERT \cite{tabert} as a transformer based encoder, makes use of database content to generate dynamic representations along with contextual encodings to represent database columns.} For a comprehensive survey covering existing solutions in NLIDB, the reader can refer to \cite{survey2019, survey2020}.

Similar to our work, Baik et. al. \cite{jagadishBridge} propose TEMPLAR, to be augmented on top of existing NLIDB systems to improve keyword mapping and therefore translation using query logs. 
\textcolor{black}{Though, TEMPLAR is not a standalone mapper, since it requires from a NLIDB system multiple preliminaries to function properly, including parsed keywords and associated metadata with each keyword, which are the main challenges yielded by the keyword mapping problem.} 
Therefore, the mapper cannot be plugged into NLIDB pipelines that does not perform detailed keyword recognition and parsing.

Different from the previous works, DBTagger is an end-to-end keyword mapper, which does not require any processing or external source of knowledge. Also, to the best of our knowledge, our work is the first study utilizing deep neural networks in a supervised learning setup for keyword mapping. 

\subsection{Deep Learning Solutions for Sequence Tagging in NLP}
In NLP community, neural network architectures have been utilized in many research problems. As a pioneer in the field, Collobert et. al. \cite{collobert2011} proposed  Convolutional Neural Networks (CNN) based architecture with CRF layer on top to deal with the sequence tagging problem. Yao et. al. \cite{Yao2014} applied LSTM in sequence tagging without having CRF as the classification layer. Bi-directional RNN structure was employed first in a speech recognition problem in \cite{bi-directionalRNN}. 

Later, instead of simple RNN networks, bi-directional LSTM was adopted and employed by Huang et. al. \cite{Huang2015BidirectionalLM} in NER problem. 
Following that study, Lample et. al. \cite{lample-etal-2016-neural} proposed a similar architecture with the inclusion of word and character embeddings. They used pre-trained word embeddings along with character level embeddings to extract input matrix to feed into the network. 
Their study stand as the state-of-the-art in sequence tagging problems in NLP. Similar to \cite{lample-etal-2016-neural}, Ma and Hovy \cite{ma-hovy-2016-end} proposed a neural architecture where character embeddings is done through CNN instead of LSTM. For a comprehensive survey discussing the deep learning solutions for research problems in NLP community, \cite{trendsNLP} is a great read.

\section{Conclusion and Future Work}
In this paper, we present DBTagger, a keyword mapper to be used in translation pipelines in NLIDB systems. DBTagger is a standalone system which does not require any processing or external knowledge such as parser or metadata preliminaries. Inspired by sequence tagging architectures used for well known problems such as POS in the NLP community, DBTagger utilizes a deep neural architecture based on bi-directional GRUs. We try to exploit the observation that POS tags of keywords are related to schema tags by applying multi-task learning in our architecture. 
DBTagger provides the best accuracy results on three publicly available databases and five schemas in Spider dataset, producing keyword tags with $92.4\%$ accuracy on the average over all the datasets within $3$ milliseconds, which is $10000$ times faster than unsupervised approaches. Our results also show that DBTagger is scalable to large databases containing millions of rows.
We believe that DBTagger can be applied in existing NLIDB systems as the first step to improve translation, especially in pipeline-based systems. For the deep learning based approaches, DBTagger can be utilized to be augmented on neural network to enrich input query before feeding into network.

\balance

\clearpage
\bibliographystyle{ACM-Reference-Format}
\bibliography{main}


\begin{thebibliography}{44}


\ifx \showCODEN    \undefined \def \showCODEN     #1{\unskip}     \fi
\ifx \showDOI      \undefined \def \showDOI       #1{#1}\fi
\ifx \showISBNx    \undefined \def \showISBNx     #1{\unskip}     \fi
\ifx \showISBNxiii \undefined \def \showISBNxiii  #1{\unskip}     \fi
\ifx \showISSN     \undefined \def \showISSN      #1{\unskip}     \fi
\ifx \showLCCN     \undefined \def \showLCCN      #1{\unskip}     \fi
\ifx \shownote     \undefined \def \shownote      #1{#1}          \fi
\ifx \showarticletitle \undefined \def \showarticletitle #1{#1}   \fi
\ifx \showURL      \undefined \def \showURL       {\relax}        \fi
\providecommand\bibfield[2]{#2}
\providecommand\bibinfo[2]{#2}
\providecommand\natexlab[1]{#1}
\providecommand\showeprint[2][]{arXiv:#2}

\bibitem[\protect\citeauthoryear{Affolter, Stockinger, and Bernstein}{Affolter
  et~al\mbox{.}}{2019}]%
        {survey2019}
\bibfield{author}{\bibinfo{person}{Katrin Affolter}, \bibinfo{person}{Kurt
  Stockinger}, {and} \bibinfo{person}{Abraham Bernstein}.}
  \bibinfo{year}{2019}\natexlab{}.
\newblock \showarticletitle{A comparative survey of recent natural language
  interfaces for databases}.
\newblock \bibinfo{journal}{\emph{The VLDB Journal}} \bibinfo{volume}{28},
  \bibinfo{number}{5} (\bibinfo{year}{2019}), \bibinfo{pages}{793--819}.
\newblock


\bibitem[\protect\citeauthoryear{{Baik}, {Jagadish}, and {Li}}{{Baik}
  et~al\mbox{.}}{2019}]%
        {jagadishBridge}
\bibfield{author}{\bibinfo{person}{Christopher {Baik}}, \bibinfo{person}{H.~V.
  {Jagadish}}, {and} \bibinfo{person}{Yunyao {Li}}.}
  \bibinfo{year}{2019}\natexlab{}.
\newblock \showarticletitle{Bridging the Semantic Gap with SQL Query Logs in
  Natural Language Interfaces to Databases}. In \bibinfo{booktitle}{\emph{2019
  IEEE 35th International Conference on Data Engineering}}
  \emph{(\bibinfo{series}{ICDE ’19})}.
\newblock


\bibitem[\protect\citeauthoryear{Basik, H\"{a}ttasch, Ilkhechi, Usta,
  Ramaswamy, Utama, Weir, Binnig, and Cetintemel}{Basik et~al\mbox{.}}{2018}]%
        {dbpal2018}
\bibfield{author}{\bibinfo{person}{Fuat Basik}, \bibinfo{person}{Benjamin
  H\"{a}ttasch}, \bibinfo{person}{Amir Ilkhechi}, \bibinfo{person}{Arif Usta},
  \bibinfo{person}{Shekar Ramaswamy}, \bibinfo{person}{Prasetya Utama},
  \bibinfo{person}{Nathaniel Weir}, \bibinfo{person}{Carsten Binnig}, {and}
  \bibinfo{person}{Ugur Cetintemel}.} \bibinfo{year}{2018}\natexlab{}.
\newblock \showarticletitle{DBPal: A Learned NL-Interface for Databases}. In
  \bibinfo{booktitle}{\emph{Proceedings of the 2018 International Conference on
  Management of Data}} \emph{(\bibinfo{series}{SIGMOD ’18})}.
  \bibinfo{pages}{1765–1768}.
\newblock


\bibitem[\protect\citeauthoryear{Blunschi, Jossen, Kossmann, Mori, and
  Stockinger}{Blunschi et~al\mbox{.}}{2012}]%
        {SODA}
\bibfield{author}{\bibinfo{person}{Lukas Blunschi}, \bibinfo{person}{Claudio
  Jossen}, \bibinfo{person}{Donald Kossmann}, \bibinfo{person}{Magdalini Mori},
  {and} \bibinfo{person}{Kurt Stockinger}.} \bibinfo{year}{2012}\natexlab{}.
\newblock \showarticletitle{SODA: Generating SQL for Business Users}.
\newblock \bibinfo{journal}{\emph{Proc. VLDB Endow.}} \bibinfo{volume}{5},
  \bibinfo{number}{10} (\bibinfo{year}{2012}).
\newblock


\bibitem[\protect\citeauthoryear{Bogin, Berant, and Gardner}{Bogin
  et~al\mbox{.}}{2019}]%
        {bogin-etal-2019-representing}
\bibfield{author}{\bibinfo{person}{Ben Bogin}, \bibinfo{person}{Jonathan
  Berant}, {and} \bibinfo{person}{Matt Gardner}.}
  \bibinfo{year}{2019}\natexlab{}.
\newblock \showarticletitle{Representing Schema Structure with Graph Neural
  Networks for Text-to-{SQL} Parsing}. In \bibinfo{booktitle}{\emph{Proceedings
  of the 57th Annual Meeting of the Association for Computational Linguistics}}
  \emph{(\bibinfo{series}{ACL ’19})}. \bibinfo{pages}{4560--4565}.
\newblock


\bibitem[\protect\citeauthoryear{Bojanowski, Grave, Joulin, and
  Mikolov}{Bojanowski et~al\mbox{.}}{2017}]%
        {fasttext}
\bibfield{author}{\bibinfo{person}{Piotr Bojanowski}, \bibinfo{person}{Edouard
  Grave}, \bibinfo{person}{Armand Joulin}, {and} \bibinfo{person}{Tomas
  Mikolov}.} \bibinfo{year}{2017}\natexlab{}.
\newblock \showarticletitle{Enriching Word Vectors with Subword Information}.
\newblock \bibinfo{journal}{\emph{Transactions of the Association for
  Computational Linguistics}}  \bibinfo{volume}{5} (\bibinfo{year}{2017}),
  \bibinfo{pages}{135--146}.
\newblock


\bibitem[\protect\citeauthoryear{Caruana}{Caruana}{1997}]%
        {caruana1997multitask}
\bibfield{author}{\bibinfo{person}{Rich Caruana}.}
  \bibinfo{year}{1997}\natexlab{}.
\newblock \showarticletitle{Multitask learning}.
\newblock \bibinfo{journal}{\emph{Machine learning}} \bibinfo{volume}{28},
  \bibinfo{number}{1} (\bibinfo{year}{1997}), \bibinfo{pages}{41--75}.
\newblock


\bibitem[\protect\citeauthoryear{Chung, Gulcehre, Cho, and Bengio}{Chung
  et~al\mbox{.}}{2015}]%
        {GRU}
\bibfield{author}{\bibinfo{person}{Junyoung Chung}, \bibinfo{person}{Caglar
  Gulcehre}, \bibinfo{person}{Kyunghyun Cho}, {and} \bibinfo{person}{Yoshua
  Bengio}.} \bibinfo{year}{2015}\natexlab{}.
\newblock \showarticletitle{Gated Feedback Recurrent Neural Networks}. In
  \bibinfo{booktitle}{\emph{Proceedings of the 32nd International Conference on
  International Conference on Machine Learning}}
  \emph{(\bibinfo{series}{ICML’15})}. \bibinfo{pages}{2067–2075}.
\newblock


\bibitem[\protect\citeauthoryear{Collobert, Weston, Bottou, Karlen,
  Kavukcuoglu, and Kuksa}{Collobert et~al\mbox{.}}{2011}]%
        {collobert2011}
\bibfield{author}{\bibinfo{person}{Ronan Collobert}, \bibinfo{person}{Jason
  Weston}, \bibinfo{person}{L\'{e}on Bottou}, \bibinfo{person}{Michael Karlen},
  \bibinfo{person}{Koray Kavukcuoglu}, {and} \bibinfo{person}{Pavel Kuksa}.}
  \bibinfo{year}{2011}\natexlab{}.
\newblock \showarticletitle{Natural Language Processing (Almost) from Scratch}.
\newblock \bibinfo{journal}{\emph{Journal of machine learning research}}
  \bibinfo{volume}{12}, \bibinfo{number}{ARTICLE} (\bibinfo{year}{2011}),
  \bibinfo{pages}{2493–2537}.
\newblock


\bibitem[\protect\citeauthoryear{Dozat}{Dozat}{2016}]%
        {nadam}
\bibfield{author}{\bibinfo{person}{Timothy Dozat}.}
  \bibinfo{year}{2016}\natexlab{}.
\newblock \showarticletitle{Incorporating Nesterov Momentum into Adam}. In
  \bibinfo{booktitle}{\emph{International Conference on Learning
  Representations Workshop}}.
\newblock


\bibitem[\protect\citeauthoryear{{Graves}, {Mohamed}, and {Hinton}}{{Graves}
  et~al\mbox{.}}{2013}]%
        {bi-directionalRNN}
\bibfield{author}{\bibinfo{person}{A. {Graves}}, \bibinfo{person}{A.
  {Mohamed}}, {and} \bibinfo{person}{G. {Hinton}}.}
  \bibinfo{year}{2013}\natexlab{}.
\newblock \showarticletitle{Speech recognition with deep recurrent neural
  networks}. In \bibinfo{booktitle}{\emph{2013 IEEE International Conference on
  Acoustics, Speech and Signal Processing}}. \bibinfo{pages}{6645--6649}.
\newblock


\bibitem[\protect\citeauthoryear{Guo, Zhan, Gao, Xiao, Lou, Liu, and Zhang}{Guo
  et~al\mbox{.}}{2019}]%
        {IRNet2019-towards}
\bibfield{author}{\bibinfo{person}{Jiaqi Guo}, \bibinfo{person}{Zecheng Zhan},
  \bibinfo{person}{Yan Gao}, \bibinfo{person}{Yan Xiao},
  \bibinfo{person}{Jian-Guang Lou}, \bibinfo{person}{Ting Liu}, {and}
  \bibinfo{person}{Dongmei Zhang}.} \bibinfo{year}{2019}\natexlab{}.
\newblock \showarticletitle{Towards Complex Text-to-{SQL} in Cross-Domain
  Database with Intermediate Representation}. In
  \bibinfo{booktitle}{\emph{Proceedings of the 57th Annual Meeting of the
  Association for Computational Linguistics}} \emph{(\bibinfo{series}{ACL
  ’19})}. \bibinfo{pages}{4524--4535}.
\newblock


\bibitem[\protect\citeauthoryear{He, Zhang, Ren, and Sun}{He
  et~al\mbox{.}}{2016}]%
        {ResNet}
\bibfield{author}{\bibinfo{person}{Kaiming He}, \bibinfo{person}{Xiangyu
  Zhang}, \bibinfo{person}{Shaoqing Ren}, {and} \bibinfo{person}{Jian Sun}.}
  \bibinfo{year}{2016}\natexlab{}.
\newblock \showarticletitle{Deep residual learning for image recognition}. In
  \bibinfo{booktitle}{\emph{Proceedings of the IEEE Conference on Computer
  Vision and Pattern Recognition}} \emph{(\bibinfo{series}{CVPR ’16})}.
  \bibinfo{pages}{770--778}.
\newblock


\bibitem[\protect\citeauthoryear{Hendrix, Sacerdoti, Sagalowicz, and
  Slocum}{Hendrix et~al\mbox{.}}{1978}]%
        {firstNLIDB}
\bibfield{author}{\bibinfo{person}{Gary~G. Hendrix}, \bibinfo{person}{Earl~D.
  Sacerdoti}, \bibinfo{person}{Daniel Sagalowicz}, {and}
  \bibinfo{person}{Jonathan Slocum}.} \bibinfo{year}{1978}\natexlab{}.
\newblock \showarticletitle{Developing a Natural Language Interface to Complex
  Data}.
\newblock \bibinfo{journal}{\emph{ACM Trans. Database Syst.}}
  \bibinfo{volume}{3}, \bibinfo{number}{2} (\bibinfo{year}{1978}),
  \bibinfo{pages}{105–147}.
\newblock


\bibitem[\protect\citeauthoryear{Hinton, Srivastava, Krizhevsky, Sutskever, and
  Salakhutdinov}{Hinton et~al\mbox{.}}{2012}]%
        {dropout}
\bibfield{author}{\bibinfo{person}{Geoffrey~E. Hinton}, \bibinfo{person}{Nitish
  Srivastava}, \bibinfo{person}{Alex Krizhevsky}, \bibinfo{person}{Ilya
  Sutskever}, {and} \bibinfo{person}{Ruslan Salakhutdinov}.}
  \bibinfo{year}{2012}\natexlab{}.
\newblock \showarticletitle{Improving neural networks by preventing
  co-adaptation of feature detectors}.
\newblock \bibinfo{journal}{\emph{ArXiv}}  \bibinfo{volume}{abs/1207.0580}
  (\bibinfo{year}{2012}).
\newblock


\bibitem[\protect\citeauthoryear{Hochreiter and Schmidhuber}{Hochreiter and
  Schmidhuber}{1997}]%
        {LSTM}
\bibfield{author}{\bibinfo{person}{Sepp Hochreiter} {and}
  \bibinfo{person}{J{\"u}rgen Schmidhuber}.} \bibinfo{year}{1997}\natexlab{}.
\newblock \showarticletitle{Long short-term memory}.
\newblock \bibinfo{journal}{\emph{Neural Computation}} \bibinfo{volume}{9},
  \bibinfo{number}{8} (\bibinfo{year}{1997}), \bibinfo{pages}{1735--1780}.
\newblock


\bibitem[\protect\citeauthoryear{Huang, Liu, Van Der~Maaten, and
  Weinberger}{Huang et~al\mbox{.}}{2017}]%
        {DenseNet}
\bibfield{author}{\bibinfo{person}{Gao Huang}, \bibinfo{person}{Zhuang Liu},
  \bibinfo{person}{Laurens Van Der~Maaten}, {and} \bibinfo{person}{Kilian~Q
  Weinberger}.} \bibinfo{year}{2017}\natexlab{}.
\newblock \showarticletitle{Densely connected convolutional networks}. In
  \bibinfo{booktitle}{\emph{Proceedings of the IEEE Conference on Computer
  Vision and Pattern Recognition}} \emph{(\bibinfo{series}{CVPR ’17})}.
  \bibinfo{pages}{4700--4708}.
\newblock


\bibitem[\protect\citeauthoryear{Huang, Wang, Singh, Yih, and He}{Huang
  et~al\mbox{.}}{2018}]%
        {huang-etal-2018-natural}
\bibfield{author}{\bibinfo{person}{Po-Sen Huang}, \bibinfo{person}{Chenglong
  Wang}, \bibinfo{person}{Rishabh Singh}, \bibinfo{person}{Wen-tau Yih}, {and}
  \bibinfo{person}{Xiaodong He}.} \bibinfo{year}{2018}\natexlab{}.
\newblock \showarticletitle{Natural Language to Structured Query Generation via
  Meta-Learning}. In \bibinfo{booktitle}{\emph{Proceedings of the 2018
  Conference of the North {A}merican Chapter of the Association for
  Computational Linguistics: Human Language Technologies, Volume 2 (Short
  Papers)}} \emph{(\bibinfo{series}{NAACL ’18})}. \bibinfo{pages}{732--738}.
\newblock


\bibitem[\protect\citeauthoryear{Huang, Xu, and Yu}{Huang
  et~al\mbox{.}}{2015}]%
        {Huang2015BidirectionalLM}
\bibfield{author}{\bibinfo{person}{Zhiheng Huang}, \bibinfo{person}{Wei Xu},
  {and} \bibinfo{person}{Kai Yu}.} \bibinfo{year}{2015}\natexlab{}.
\newblock \showarticletitle{Bidirectional LSTM-CRF Models for Sequence
  Tagging}.
\newblock \bibinfo{journal}{\emph{ArXiv}}  \bibinfo{volume}{abs/1508.01991}
  (\bibinfo{year}{2015}).
\newblock


\bibitem[\protect\citeauthoryear{Iyer, Konstas, Cheung, Krishnamurthy, and
  Zettlemoyer}{Iyer et~al\mbox{.}}{2017}]%
        {iyer-etal-2017-learning}
\bibfield{author}{\bibinfo{person}{Srinivasan Iyer}, \bibinfo{person}{Ioannis
  Konstas}, \bibinfo{person}{Alvin Cheung}, \bibinfo{person}{Jayant
  Krishnamurthy}, {and} \bibinfo{person}{Luke Zettlemoyer}.}
  \bibinfo{year}{2017}\natexlab{}.
\newblock \showarticletitle{Learning a Neural Semantic Parser from User
  Feedback}. In \bibinfo{booktitle}{\emph{Proceedings of the 55th Annual
  Meeting of the Association for Computational Linguistics}}
  \emph{(\bibinfo{series}{ACL ’17})}. \bibinfo{pages}{963--973}.
\newblock


\bibitem[\protect\citeauthoryear{Kim, So, Han, and Lee}{Kim
  et~al\mbox{.}}{2020}]%
        {survey2020}
\bibfield{author}{\bibinfo{person}{Hyeonji Kim}, \bibinfo{person}{Byeong-Hoon
  So}, \bibinfo{person}{Wook-Shin Han}, {and} \bibinfo{person}{Hongrae Lee}.}
  \bibinfo{year}{2020}\natexlab{}.
\newblock \showarticletitle{Natural language to SQL: Where are we today?}
\newblock \bibinfo{journal}{\emph{Proceedings of the VLDB Endowment}}
  \bibinfo{volume}{13}, \bibinfo{number}{10} (\bibinfo{year}{2020}),
  \bibinfo{pages}{1737--1750}.
\newblock


\bibitem[\protect\citeauthoryear{Lafferty, McCallum, and Pereira}{Lafferty
  et~al\mbox{.}}{2001}]%
        {CRF}
\bibfield{author}{\bibinfo{person}{John~D. Lafferty}, \bibinfo{person}{Andrew
  McCallum}, {and} \bibinfo{person}{Fernando C.~N. Pereira}.}
  \bibinfo{year}{2001}\natexlab{}.
\newblock \showarticletitle{Conditional Random Fields: Probabilistic Models for
  Segmenting and Labeling Sequence Data}. In
  \bibinfo{booktitle}{\emph{Proceedings of the Eighteenth International
  Conference on Machine Learning}} \emph{(\bibinfo{series}{ICML ’01})}.
  \bibinfo{pages}{282–289}.
\newblock


\bibitem[\protect\citeauthoryear{Lample, Ballesteros, Subramanian, Kawakami,
  and Dyer}{Lample et~al\mbox{.}}{2016}]%
        {lample-etal-2016-neural}
\bibfield{author}{\bibinfo{person}{Guillaume Lample}, \bibinfo{person}{Miguel
  Ballesteros}, \bibinfo{person}{Sandeep Subramanian}, \bibinfo{person}{Kazuya
  Kawakami}, {and} \bibinfo{person}{Chris Dyer}.}
  \bibinfo{year}{2016}\natexlab{}.
\newblock \showarticletitle{Neural Architectures for Named Entity Recognition}.
  In \bibinfo{booktitle}{\emph{Proceedings of the 2016 Conference of the North
  {A}merican Chapter of the Association for Computational Linguistics: Human
  Language Technologies}} \emph{(\bibinfo{series}{NAACL ’16})}.
  \bibinfo{pages}{260--270}.
\newblock


\bibitem[\protect\citeauthoryear{Li and Jagadish}{Li and Jagadish}{2014}]%
        {NALIR}
\bibfield{author}{\bibinfo{person}{Fei Li} {and} \bibinfo{person}{H.~V.
  Jagadish}.} \bibinfo{year}{2014}\natexlab{}.
\newblock \showarticletitle{Constructing an Interactive Natural Language
  Interface for Relational Databases}.
\newblock \bibinfo{journal}{\emph{Proc. VLDB Endow.}} \bibinfo{volume}{8},
  \bibinfo{number}{1} (\bibinfo{year}{2014}), \bibinfo{pages}{73–84}.
\newblock


\bibitem[\protect\citeauthoryear{Li and Rafiei}{Li and Rafiei}{2017}]%
        {keywordChallenge}
\bibfield{author}{\bibinfo{person}{Yunyao Li} {and} \bibinfo{person}{Davood
  Rafiei}.} \bibinfo{year}{2017}\natexlab{}.
\newblock \showarticletitle{Natural Language Data Management and Interfaces:
  Recent Development and Open Challenges}. In
  \bibinfo{booktitle}{\emph{Proceedings of the 2017 ACM International
  Conference on Management of Data}} \emph{(\bibinfo{series}{SIGMOD ’17})}.
  \bibinfo{pages}{1765–1770}.
\newblock


\bibitem[\protect\citeauthoryear{Ma and Hovy}{Ma and Hovy}{2016}]%
        {ma-hovy-2016-end}
\bibfield{author}{\bibinfo{person}{Xuezhe Ma} {and} \bibinfo{person}{Eduard
  Hovy}.} \bibinfo{year}{2016}\natexlab{}.
\newblock \showarticletitle{End-to-end Sequence Labeling via Bi-directional
  {LSTM}-{CNN}s-{CRF}}. In \bibinfo{booktitle}{\emph{Proceedings of the 54th
  Annual Meeting of the Association for Computational Linguistics}}
  \emph{(\bibinfo{series}{ACL ’16})}. \bibinfo{pages}{1064--1074}.
\newblock


\bibitem[\protect\citeauthoryear{Manning, Surdeanu, Bauer, Finkel, Bethard, and
  McClosky}{Manning et~al\mbox{.}}{2014}]%
        {coreNLP}
\bibfield{author}{\bibinfo{person}{Christopher~D. Manning},
  \bibinfo{person}{Mihai Surdeanu}, \bibinfo{person}{John Bauer},
  \bibinfo{person}{Jenny Finkel}, \bibinfo{person}{Steven~J. Bethard}, {and}
  \bibinfo{person}{David McClosky}.} \bibinfo{year}{2014}\natexlab{}.
\newblock \showarticletitle{The {Stanford} {CoreNLP} Natural Language
  Processing Toolkit}. In \bibinfo{booktitle}{\emph{Association for
  Computational Linguistics (ACL) System Demonstrations}}.
  \bibinfo{pages}{55--60}.
\newblock


\bibitem[\protect\citeauthoryear{Mikolov, Corrado, Chen, and Dean}{Mikolov
  et~al\mbox{.}}{2013}]%
        {word2vec}
\bibfield{author}{\bibinfo{person}{Tomas Mikolov}, \bibinfo{person}{G.s
  Corrado}, \bibinfo{person}{Kai Chen}, {and} \bibinfo{person}{Jeffrey Dean}.}
  \bibinfo{year}{2013}\natexlab{}.
\newblock \showarticletitle{Efficient Estimation of Word Representations in
  Vector Space}. In \bibinfo{booktitle}{\emph{Proceedings of the International
  Conference on Learning Representations}} \emph{(\bibinfo{series}{ICLR'13})}.
  \bibinfo{pages}{1--12}.
\newblock


\bibitem[\protect\citeauthoryear{Miller}{Miller}{1995}]%
        {wordNet}
\bibfield{author}{\bibinfo{person}{George~A. Miller}.}
  \bibinfo{year}{1995}\natexlab{}.
\newblock \showarticletitle{WordNet: A Lexical Database for English}.
\newblock \bibinfo{journal}{\emph{Commun. ACM}} \bibinfo{volume}{38},
  \bibinfo{number}{11} (\bibinfo{year}{1995}), \bibinfo{pages}{39–41}.
\newblock


\bibitem[\protect\citeauthoryear{\"{O}zcan, Quamar, Sen, Lei, and
  Efthymiou}{\"{O}zcan et~al\mbox{.}}{2020}]%
        {challenges2020}
\bibfield{author}{\bibinfo{person}{Fatma \"{O}zcan}, \bibinfo{person}{Abdul
  Quamar}, \bibinfo{person}{Jaydeep Sen}, \bibinfo{person}{Chuan Lei}, {and}
  \bibinfo{person}{Vasilis Efthymiou}.} \bibinfo{year}{2020}\natexlab{}.
\newblock \showarticletitle{State of the Art and Open Challenges in Natural
  Language Interfaces to Data}. In \bibinfo{booktitle}{\emph{Proceedings of the
  2020 ACM SIGMOD International Conference on Management of Data}}
  \emph{(\bibinfo{series}{SIGMOD ’20})}. \bibinfo{pages}{2629–2636}.
\newblock


\bibitem[\protect\citeauthoryear{Popescu, Etzioni, and Kautz}{Popescu
  et~al\mbox{.}}{2003}]%
        {PrecisePopescu}
\bibfield{author}{\bibinfo{person}{Ana-Maria Popescu}, \bibinfo{person}{Oren
  Etzioni}, {and} \bibinfo{person}{Henry Kautz}.}
  \bibinfo{year}{2003}\natexlab{}.
\newblock \showarticletitle{Towards a Theory of Natural Language Interfaces to
  Databases}. In \bibinfo{booktitle}{\emph{Proceedings of the 8th International
  Conference on Intelligent User Interfaces}} \emph{(\bibinfo{series}{IUI
  ’03})}. \bibinfo{pages}{149–157}.
\newblock


\bibitem[\protect\citeauthoryear{Saha, Floratou, Sankaranarayanan, Minhas,
  Mittal, and \"{O}zcan}{Saha et~al\mbox{.}}{2016}]%
        {ATHENA}
\bibfield{author}{\bibinfo{person}{Diptikalyan Saha}, \bibinfo{person}{Avrilia
  Floratou}, \bibinfo{person}{Karthik Sankaranarayanan},
  \bibinfo{person}{Umar~Farooq Minhas}, \bibinfo{person}{Ashish~R. Mittal},
  {and} \bibinfo{person}{Fatma \"{O}zcan}.} \bibinfo{year}{2016}\natexlab{}.
\newblock \showarticletitle{ATHENA: An Ontology-Driven System for Natural
  Language Querying over Relational Data Stores}.
\newblock \bibinfo{journal}{\emph{Proc. VLDB Endow.}} \bibinfo{volume}{9},
  \bibinfo{number}{12} (\bibinfo{year}{2016}).
\newblock


\bibitem[\protect\citeauthoryear{Weir, Utama, Galakatos, Crotty, Ilkhechi,
  Ramaswamy, Bhushan, Geisler, H\"{a}ttasch, Eger, Cetintemel, and Binnig}{Weir
  et~al\mbox{.}}{2020}]%
        {dbpal2020}
\bibfield{author}{\bibinfo{person}{Nathaniel Weir}, \bibinfo{person}{Prasetya
  Utama}, \bibinfo{person}{Alex Galakatos}, \bibinfo{person}{Andrew Crotty},
  \bibinfo{person}{Amir Ilkhechi}, \bibinfo{person}{Shekar Ramaswamy},
  \bibinfo{person}{Rohin Bhushan}, \bibinfo{person}{Nadja Geisler},
  \bibinfo{person}{Benjamin H\"{a}ttasch}, \bibinfo{person}{Steffen Eger},
  \bibinfo{person}{Ugur Cetintemel}, {and} \bibinfo{person}{Carsten Binnig}.}
  \bibinfo{year}{2020}\natexlab{}.
\newblock \showarticletitle{DBPal: A Fully Pluggable NL2SQL Training Pipeline}.
  In \bibinfo{booktitle}{\emph{Proceedings of the 2020 ACM SIGMOD International
  Conference on Management of Data}} \emph{(\bibinfo{series}{SIGMOD ’20})}.
  \bibinfo{pages}{2347–2361}.
\newblock


\bibitem[\protect\citeauthoryear{Xu, Liu, and Song}{Xu et~al\mbox{.}}{2017}]%
        {xu2017sqlnet}
\bibfield{author}{\bibinfo{person}{Xiaojun Xu}, \bibinfo{person}{Chang Liu},
  {and} \bibinfo{person}{Dawn Song}.} \bibinfo{year}{2017}\natexlab{}.
\newblock \showarticletitle{Sqlnet: Generating structured queries from natural
  language without reinforcement learning}.
\newblock \bibinfo{journal}{\emph{arXiv preprint arXiv:1711.04436}}
  (\bibinfo{year}{2017}).
\newblock


\bibitem[\protect\citeauthoryear{Yaghmazadeh, Wang, Dillig, and
  Dillig}{Yaghmazadeh et~al\mbox{.}}{2017}]%
        {Sqlizer}
\bibfield{author}{\bibinfo{person}{Navid Yaghmazadeh}, \bibinfo{person}{Yuepeng
  Wang}, \bibinfo{person}{Isil Dillig}, {and} \bibinfo{person}{Thomas Dillig}.}
  \bibinfo{year}{2017}\natexlab{}.
\newblock \showarticletitle{SQLizer: Query Synthesis from Natural Language}.
\newblock \bibinfo{journal}{\emph{Proceedings of the ACM on Programming
  Languages}} \bibinfo{volume}{1}, \bibinfo{number}{OOPSLA}
  (\bibinfo{year}{2017}), \bibinfo{pages}{1--26}.
\newblock


\bibitem[\protect\citeauthoryear{Yao, Peng, Zhang, Yu, Zweig, and Shi}{Yao
  et~al\mbox{.}}{2014}]%
        {Yao2014}
\bibfield{author}{\bibinfo{person}{Kaisheng Yao}, \bibinfo{person}{Baolin
  Peng}, \bibinfo{person}{Yu Zhang}, \bibinfo{person}{Dong Yu},
  \bibinfo{person}{Geoffrey Zweig}, {and} \bibinfo{person}{Yangyang Shi}.}
  \bibinfo{year}{2014}\natexlab{}.
\newblock \showarticletitle{Spoken language understanding using long short-term
  memory neural networks}. In \bibinfo{booktitle}{\emph{2014 IEEE Spoken
  Language Technology Workshop (SLT)}}. \bibinfo{pages}{189--194}.
\newblock


\bibitem[\protect\citeauthoryear{Yavuz, Gur, Su, and Yan}{Yavuz
  et~al\mbox{.}}{2018}]%
        {yavuz-etal-2018-takes}
\bibfield{author}{\bibinfo{person}{Semih Yavuz}, \bibinfo{person}{Izzeddin
  Gur}, \bibinfo{person}{Yu Su}, {and} \bibinfo{person}{Xifeng Yan}.}
  \bibinfo{year}{2018}\natexlab{}.
\newblock \showarticletitle{What It Takes to Achieve 100{\%} Condition Accuracy
  on {W}iki{SQL}}. In \bibinfo{booktitle}{\emph{Proceedings of the 2018
  Conference on Empirical Methods in Natural Language Processing}}
  \emph{(\bibinfo{series}{EMNLP ’18})}. \bibinfo{pages}{1702--1711}.
\newblock


\bibitem[\protect\citeauthoryear{Yin, Neubig, Yih, and Riedel}{Yin
  et~al\mbox{.}}{2020}]%
        {tabert}
\bibfield{author}{\bibinfo{person}{Pengcheng Yin}, \bibinfo{person}{Graham
  Neubig}, \bibinfo{person}{Wen-tau Yih}, {and} \bibinfo{person}{Sebastian
  Riedel}.} \bibinfo{year}{2020}\natexlab{}.
\newblock \showarticletitle{{T}a{BERT}: Pretraining for Joint Understanding of
  Textual and Tabular Data}. In \bibinfo{booktitle}{\emph{Proceedings of the
  58th Annual Meeting of the Association for Computational Linguistics}}
  \emph{(\bibinfo{series}{ACL ’20})}. \bibinfo{pages}{8413--8426}.
\newblock


\bibitem[\protect\citeauthoryear{Young, Hazarika, Poria, and Cambria}{Young
  et~al\mbox{.}}{2018}]%
        {trendsNLP}
\bibfield{author}{\bibinfo{person}{Tom Young}, \bibinfo{person}{Devamanyu
  Hazarika}, \bibinfo{person}{Soujanya Poria}, {and} \bibinfo{person}{Erik
  Cambria}.} \bibinfo{year}{2018}\natexlab{}.
\newblock \showarticletitle{Recent trends in deep learning based natural
  language processing}.
\newblock \bibinfo{journal}{\emph{IEEE Computational Intelligence Magazine}}
  \bibinfo{volume}{13}, \bibinfo{number}{3} (\bibinfo{year}{2018}),
  \bibinfo{pages}{55--75}.
\newblock


\bibitem[\protect\citeauthoryear{Yu, Li, Zhang, Zhang, and Radev}{Yu
  et~al\mbox{.}}{2018a}]%
        {yu-etal-2018-typesql}
\bibfield{author}{\bibinfo{person}{Tao Yu}, \bibinfo{person}{Zifan Li},
  \bibinfo{person}{Zilin Zhang}, \bibinfo{person}{Rui Zhang}, {and}
  \bibinfo{person}{Dragomir Radev}.} \bibinfo{year}{2018}\natexlab{a}.
\newblock \showarticletitle{{T}ype{SQL}: Knowledge-Based Type-Aware Neural
  Text-to-{SQL} Generation}. In \bibinfo{booktitle}{\emph{Proceedings of the
  2018 Conference of the North {A}merican Chapter of the Association for
  Computational Linguistics: Human Language Technologies, Volume 2 (Short
  Papers)}} \emph{(\bibinfo{series}{NAACL ’18})}. \bibinfo{pages}{588--594}.
\newblock


\bibitem[\protect\citeauthoryear{Yu, Yasunaga, Yang, Zhang, Wang, Li, and
  Radev}{Yu et~al\mbox{.}}{2018b}]%
        {syntaxSQL}
\bibfield{author}{\bibinfo{person}{Tao Yu}, \bibinfo{person}{Michihiro
  Yasunaga}, \bibinfo{person}{Kai Yang}, \bibinfo{person}{Rui Zhang},
  \bibinfo{person}{Dongxu Wang}, \bibinfo{person}{Zifan Li}, {and}
  \bibinfo{person}{Dragomir Radev}.} \bibinfo{year}{2018}\natexlab{b}.
\newblock \showarticletitle{{S}yntax{SQLN}et: Syntax Tree Networks for Complex
  and Cross-Domain Text-to-{SQL} Task}. In
  \bibinfo{booktitle}{\emph{Proceedings of the 2018 Conference on Empirical
  Methods in Natural Language Processing}} \emph{(\bibinfo{series}{EMNLP
  ’18})}. \bibinfo{pages}{1653--1663}.
\newblock


\bibitem[\protect\citeauthoryear{Yu, Zhang, Yang, Yasunaga, Wang, Li, Ma, Li,
  Yao, Roman, Zhang, and Radev}{Yu et~al\mbox{.}}{2018c}]%
        {yu-etal-2018-spider}
\bibfield{author}{\bibinfo{person}{Tao Yu}, \bibinfo{person}{Rui Zhang},
  \bibinfo{person}{Kai Yang}, \bibinfo{person}{Michihiro Yasunaga},
  \bibinfo{person}{Dongxu Wang}, \bibinfo{person}{Zifan Li},
  \bibinfo{person}{James Ma}, \bibinfo{person}{Irene Li},
  \bibinfo{person}{Qingning Yao}, \bibinfo{person}{Shanelle Roman},
  \bibinfo{person}{Zilin Zhang}, {and} \bibinfo{person}{Dragomir Radev}.}
  \bibinfo{year}{2018}\natexlab{c}.
\newblock \showarticletitle{{S}pider: A Large-Scale Human-Labeled Dataset for
  Complex and Cross-Domain Semantic Parsing and Text-to-{SQL} Task}. In
  \bibinfo{booktitle}{\emph{Proceedings of the 2018 Conference on Empirical
  Methods in Natural Language Processing}} \emph{(\bibinfo{series}{EMNLP
  ’18})}. \bibinfo{pages}{3911--3921}.
\newblock


\bibitem[\protect\citeauthoryear{Zeiler}{Zeiler}{2012}]%
        {Zeiler2012ADADELTAAA}
\bibfield{author}{\bibinfo{person}{Matthew~D. Zeiler}.}
  \bibinfo{year}{2012}\natexlab{}.
\newblock \showarticletitle{ADADELTA: An Adaptive Learning Rate Method}.
\newblock \bibinfo{journal}{\emph{ArXiv}}  \bibinfo{volume}{abs/1212.5701}
  (\bibinfo{year}{2012}).
\newblock


\bibitem[\protect\citeauthoryear{Zhong, Xiong, and Socher}{Zhong
  et~al\mbox{.}}{2017}]%
        {zhong2017seq2sql}
\bibfield{author}{\bibinfo{person}{Victor Zhong}, \bibinfo{person}{Caiming
  Xiong}, {and} \bibinfo{person}{Richard Socher}.}
  \bibinfo{year}{2017}\natexlab{}.
\newblock \showarticletitle{Seq2sql: Generating structured queries from natural
  language using reinforcement learning}.
\newblock \bibinfo{journal}{\emph{arXiv preprint arXiv:1709.00103}}
  (\bibinfo{year}{2017}).
\newblock


\end{thebibliography}

\end{document}